\documentclass{PoS}
\pdfoutput=1

\usepackage{amsmath}

\usepackage[numbers,sort&compress]{natbib}
\usepackage{natbibspacing}
\newcommand{\matrixel}[3]{\left< #1 \vphantom{#2#3} \right| #2 \left| #3 \vphantom{#1#2} \right>}
\newcommand{\abs}[1]{\left| #1 \right|}
\newcommand{\avg}[1]{\left< #1 \right>}

\newcommand{\eq}[1]{\begin{equation} #1 \end{equation}}
\newcommand{\al}[1]{\begin{equation} \begin{aligned} #1 \end{aligned} \end{equation}}
\newcommand{\mc}[1]{\mathcal{#1}} 
\usepackage{wasysym}

\title{Charm Physics with Domain Wall Fermions and Physical Pion Masses}

\ShortTitle{Domain Wall Charm Physics with Physical Pion Masses}

\author{Peter Boyle$^a$, Luigi Del Debbio$^a$, Andreas J\"uttner$^b$, Ava Khamseh$^a$, Francesco Sanfilippo$^b$, \speaker{Justus Tobias Tsang}$^{\,\,a,b}$, Oliver Witzel$^a$\\
  \llap{$^a$} School of Physics and Astronomy, University of Edinburgh\\
  EH9 3JZ, Edinburgh, United Kingdom\\
  \llap{$^b$} School of Physics and Astronomy, University of Southampton\\
  SO17 1BJ Southampton, United Kingdom\\
  E-mail: \email{j.t.tsang@ed.ac.uk}
}

\author{RBC and UKQCD Collaborations}

\abstract{We present RBC/UKQCD's charm project using $N_f=2+1$ flavour ensembles with inverse lattice spacings in the range $1.73-2.77\,\mathrm{GeV}$ and two physical pion mass ensembles. Domain wall fermions are used for the light as well as the charm quarks. We discuss our strategy for the extraction of the decay constants $f_D$ and $f_{D_s}$ and their extrapolation to the continuum limit, physical pion masses and the physical heavy quark mass. Our preliminary results are $f_D=208.7(2.8)\,\mathrm{MeV}$ and $f_{D_s}=246.4(1.9)\,\mathrm{MeV}$ where the quoted error is statistical only.
We outline our current approach to extend the reach in the heavy quark mass and present preliminary results.}

\FullConference{34th annual International Symposium on Lattice Field Theory\\
		24-30 July 2016\\
		University of Southampton, UK}

\begin{document}
%%%%%%%%%%%%%%%%%%%%%%%%%%%%%%%%%%%%%%%%%5
\section{Introduction}
%%%%%%%%%%%%%%%%%%%%%%%%%%%%%%%%%%%%%%%%%5
In this work we present preliminary results of RBC/UKQCD's $D$ and $D_s$ mesons decay constant studies using $N_f=2+1$ flavour gauge field ensembles with domain wall fermions, updating refs ~\cite{Boyle:2015rka,Boyle:2015kyy}.
We aim to predict the $D$ and $D_s$ decay constants with a fully controlled systematic error budget. The decay constant $f_{D_q}$ of the $D_{(s)}$ meson is defined as
\eq{
  \matrixel{0}{A_{cq}^\mu}{D_q(p)} = f_{D_q} p^\mu_{D_q},
  \label{eq:decayconst}
}
where $q=d,s$ and the axial current is defined as $A_{cq}^\mu = \overline{c}\gamma_\mu \gamma_5 q$. 
Combined with experimental measurements of the decay widths $\Gamma \left(D_{(s)} \to l\nu_l\right)$ knowledge of the decay constants allows to extract the CKM matrix elements $\abs{V_{cd}}$ and $\abs{V_{cs}}$ and hence test the unitarity of the CKM matrix.

For this we use seven ensembles specified in Tables \ref{tab:ensembles} and \ref{tab:light_strange} with the Iwasaki gauge action~\cite{Iwasakiaction2} and the domain wall fermion action~\cite{Kaplan:1992bt,Shamir:1993} using the Moebius-kernel~\cite{Brower:Mobius}. We use ensembles with three different lattice spacings in the range $a^{-1}=1.73-2.77\,\mathrm{GeV}$, including two ensembles with near physical pion masses~\cite{PhysicalPoint}. 

After a brief summary of our ensembles and set-up in Section \ref{sec:ensembles} we will outline our analysis strategy to obtain predictions for physical masses in the continuum limit and give preliminary results in Section \ref{sec:analysis}.

Finally, we will outline the status of current projects and conclude in Section \ref{sec:conclusions}.

%%%%%%%%%%%%%%%%%%%%%%%%%%%%%%%%%%%%%%%%%5
\section{Ensembles and Run Set-Up} \label{sec:ensembles}
%%%%%%%%%%%%%%%%%%%%%%%%%%%%%%%%%%%%%%%%%5
\begin{table}
  \begin{center}
    \begin{tabular}{|cc|ccccc|cc|}
      \hline
      $\beta$ & Name & $L^3 \times T /a^4$ &  $a^{-1}[\mathrm{GeV}]$ & $L\,[\mathrm{fm}]$ & $m_\pi[\mathrm{MeV}$] & $m_\pi L $  & $N_{\mathrm{conf}}$ & $N_\mathrm{meas}$\\\hline
      2.13 & C0  & $48^3 \times 96$    & 1.7295(38) & 5.4760(24) & {\bf 139}  & 3.863(05) & 88& 4224\\
      2.13 & C1  & $24^3 \times 64$    & 1.7848(50) & 2.6532(15) & 340  & 4.570(10) &100& 3200\\
      2.13 & C2  & $24^3 \times 64$    & 1.7848(50) & 2.6532(15) & 430  & 5.790(10) &101& 3232\\\hline
      2.25 & M0  & $64^3 \times 128$   & 2.3586(70) & 5.3539(31) & {\bf 139}  & 3.781(05) &80& 2560\\
      2.25 & M1  & $32^3 \times 64$    & 2.3833(86) & 2.6492(19) & 300  & 4.072(11) &83& 2656 \\
      2.25 & M2  & $32^3 \times 64$    & 2.3833(86) & 2.6492(19) & 360  & 4.854(12) &76& 1216\\\hline
      2.31 & F1  & $48^3 \times 96$    & 2.774(10)$\hphantom{0}$  & 3.4141(24) & 235  & 4.053(06) &82& 3936\\
      \hline 
    \end{tabular}
  \end{center}
  \caption{This table summarises the main parameters of the used ensembles. All of these use the Iwasaki gauge action with $2+1$ flavours of domain wall fermions in the sea. $N_\mathrm{conf}$ and $N_\mathrm{meas}$ refer to the number of (decorrelated) configurations used and the number of total measurements, respectively.}
  \label{tab:ensembles}
\end{table}

Table \ref{tab:ensembles} provides a brief summary of the ensembles used for this study.
The sea quarks of all ensembles are Moebius domain wall fermions~\cite{Brower:Mobius}. The ensembles C1, C2, M1 and M2 use the Moebius scale factor $\alpha=1$ which reproduces Shamir domain wall fermions~\cite{Shamir:1993}, whilst C0, M0 and F1 have been generated with $\alpha=2$ ensuring the same approach to the continuum, as outlined in ref \cite{PhysicalPoint}.

The domain wall parameters of the sea quarks (i.e. the domain wall height $M_5$ and extent of the fifth dimension $L_s$) are listed in Table \ref{tab:light_strange}.
We keep these same values for the valence light and strange quarks. 
The physical value of the strange quark mass in bare lattice units is determined by reproducing the procedure of ref \cite{PhysicalPoint} including the new ensemble F1 and the results are listed in Table \ref{tab:light_strange}. 
Whilst the light quark mass is always kept unitary, we adjust the valence strange quark mass to its physical value (compare Table \ref{tab:light_strange}) on measurements on the ensembles C1, C2, M1 and M2. On the remaining ensembles the valence strange quark mass is kept to be the unitary one, forcing a small correction to its physical value, which will be discussed below.

\begin{table}
  \begin{center}
    \begin{tabular}{|c||l|ll|l||r|l|}
      \hline
      & \multicolumn{4}{c||}{Moebius/Shamir with $M_5=1.8$} & \multicolumn{2}{c|}{Moebius with $M_5=1.6$} \\\hline
      Name & $L_s$ & $am_l^\mathrm{sea}$ &  $am_s^\mathrm{sea}$ & $am_s^\mathrm{phys}$ & $L_s$ & $am_h$\\\hline
      C0 & 24 & 0.00078  & 0.0362  & 0.03580(16) & 12 & 0.30, 0.35, 0.40 \\
      C1 & 16 & 0.005    & 0.04    & 0.03224(18) & 12 & 0.30, 0.35, 0.40 \\
      C2 & 16 & 0.01     & 0.04    & 0.03224(18) & 12 & 0.30, 0.35, 0.40 \\\hline
      M0 & 12 & 0.000678 & 0.02661 & 0.02540(17) &  8 & 0.22, 0.28, 0.34, 0.40 \\
      M1 & 16 & 0.004    & 0.03    & 0.02477(18) & 12 & 0.22, 0.28, 0.34, 0.40\\
      M2 & 16 & 0.006    & 0.03    & 0.02477(18) & 12 & 0.22, 0.28, 0.34, 0.40\\\hline
      F1 & 12 & 0.002144 & 0.02144 & 0.02132(17) & 12 & 0.18, 0.23, 0.28, 0.33, 0.40\\
      \hline 
    \end{tabular}
  \end{center}
  \caption{Ensemble and measurement parameters for the propagators used on RBC/UKQCD's $N_f=2+1$ flavour Iwasaki gauge action ensembles. The choice of $(M_5,L_s)$ differs between the light and strange (left) and the heavy (right) propagators, leading to a mixed action.}
  \label{tab:light_strange}
\end{table}

In ref \cite{Boyle:2016imm} we studied the Moebius domain wall parameter space with the intention of finding a region in which discretisation effects for charmed meson observables are small.
We found that the choice of the domain wall height $M_5=1.6$ yields optimal results, provided the bound $am^\mathrm{bare}_h \lesssim 0.4$ is observed.
In this work we adopt these choices for the charm quark. We note that this leads to a mixed action since the two quarks entering the current have a different discretisation due to the different choice of $M_5$. However, we show effects from this are small~\cite{dynamical2016}.

We use stochastic $\mathbb{Z}_2\times\mathbb{Z}_2$-sources (Z2-Wall)~\cite{Foster:1998vw} on every fourth ($\Delta T=4$) time plane for M0 and M2 and every second ($\Delta T=2$) time plane for all other ensembles, yielding a stochastic estimate ($L^3 \times T/\Delta T$) for the full volume average of the correlation functions. The detailed statistics are listed in Table \ref{tab:ensembles}. Light and strange quark propagators have been computed with the HDCG~\cite{HDCG} algorithm, whilst for the heavy quark propagators a CG inverter was used. 

%%%%%%%%%%%%%%%%%%%%%%%%%%%%%%%%%%%%%%%%%5
\section{Decay Constant Analysis} \label{sec:analysis}
\begin{figure}
  \begin{center}
    \includegraphics[width=0.45\textwidth]{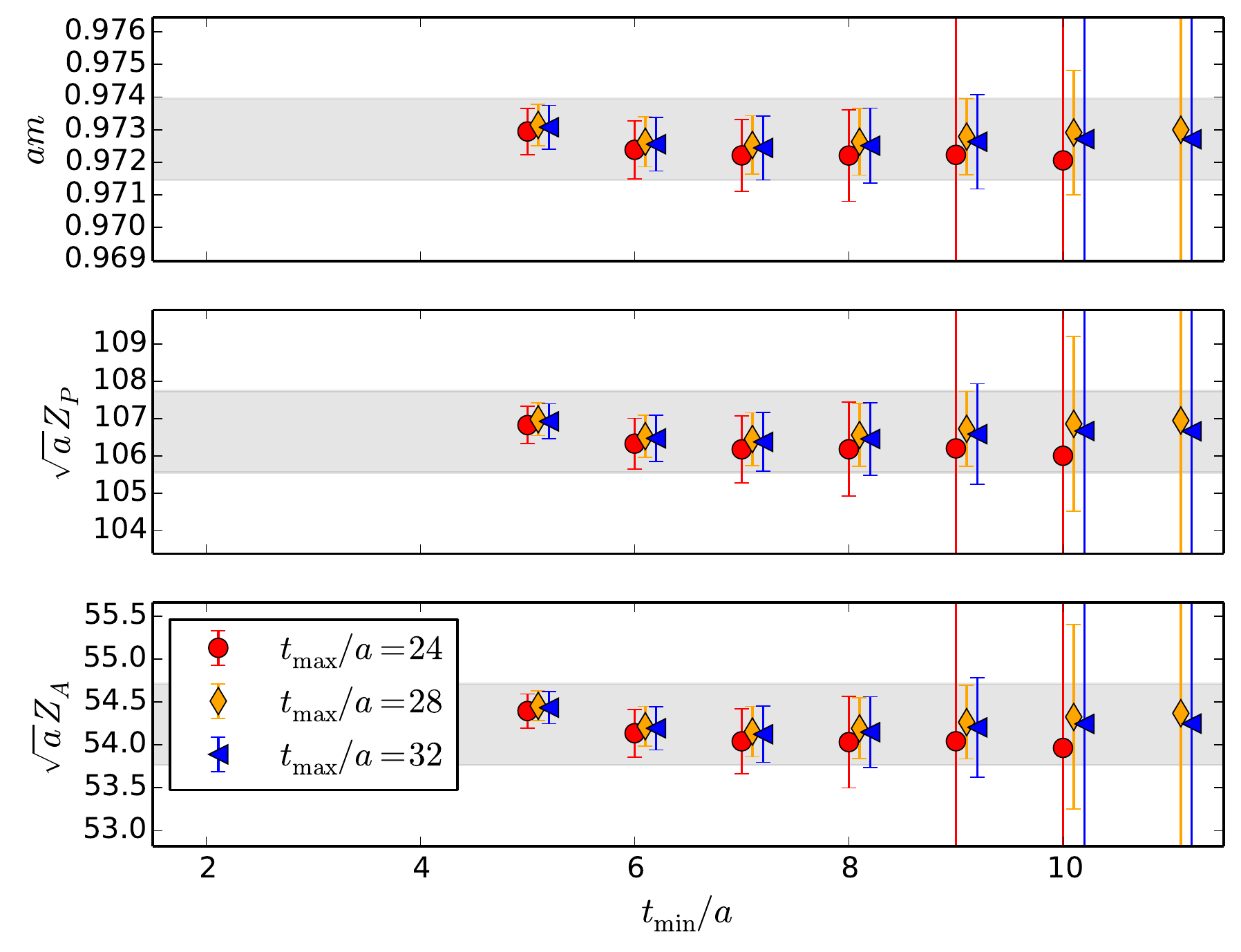}
    \includegraphics[width=0.45\textwidth]{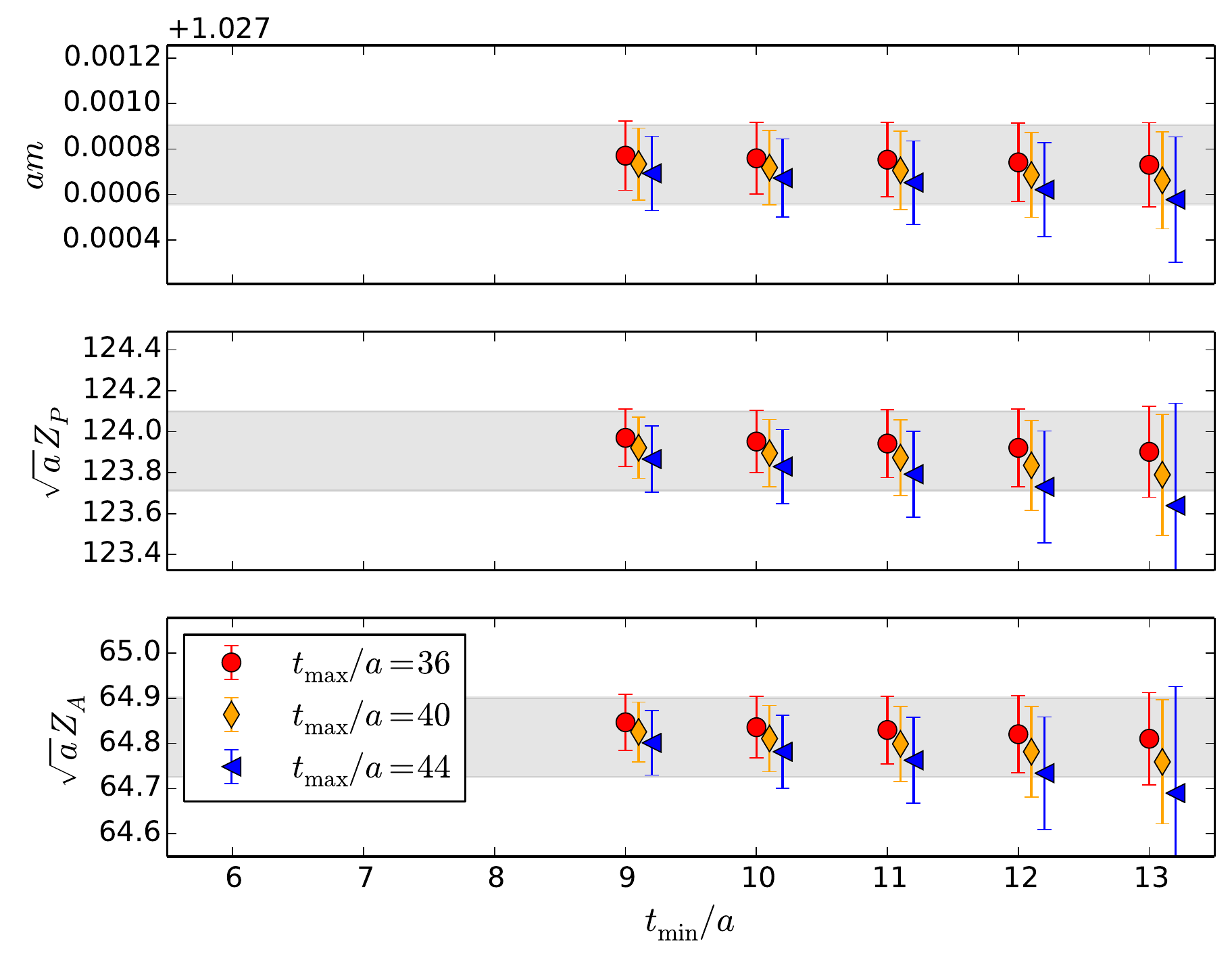}
  \end{center}
  \caption{Variation of fit ranges illustrated for the heaviest mass point of the ensemble C0 for a light-heavy (left) and a strange-heavy (right) meson. The three panels correspond to fit results of the ground state. The $x$-axis shows the initial time slice $t_\mathrm{min}$ entering the fit, whilst the choice of symbol and colour indicates the final time slice $t_\mathrm{max}$ included in the fit. The grey band shows the (conservatively) chosen fit for the subsequent analysis with fit ranges $[t_\mathrm{min},t_\mathrm{max})=[8,30)$ (light-heavy) and $[12,37)$ (strange-heavy), respectively.}
  \label{fig:fitranges}
\end{figure}

%%%%%%%%%%%%%%%%%%%%%%%%%%%%%%%%%%%%%%%%%5
To determine the decay constant we are interested in the matrix element $\matrixel{0}{A_{cq}^4}{D_q({\bf p}={\bf 0})}$ (compare eq \eqref{eq:decayconst}).
Masses and matrix elements can be extracted from fits to correlation functions. 
We simultaneously fitted the ground state and the first excited state of the meson two-point functions $\avg{AP}$ and $\avg{PP}$ where $A$ is the operator for the temporal component of the axial current and $P$ is the pseudoscalar operator. We studied the impact of the choice of fit ranges by systematically varying the initial ($t_\mathrm{min}$) and final ($t_\mathrm{max}$) time slices entering the fit and conservatively chose these values as shown in Figure \ref{fig:fitranges}.

From the matrix elements we can extract the decay constants $f_P$ with $P=D,D_{s}$ and their ratio.
The decay amplitude $\Phi_P = f_P\sqrt{m_P}$ displays a nearly linear behaviour with the inverse heavy quark mass, so for convenience we carry out the remaining analysis in terms of this quantity. 
The following procedure is two-fold.
We first renormalise the decay constants and correct for the small mistuning in the strange quark mass mentioned above.
We then carry out a combined fit to obtain physical heavy and light quark masses (by using appropriate meson masses) and remove lattice artifacts.

\subsection{Renormalisation and strange quark mistuning}
Due to the mixed action approach, we do not have a conserved heavy-light current.
Instead we determine the renormalisation constant $\mc{Z}_A$ from the ratio of local and conserved light-light currents and renormalise the data using this.
From an NPR study we find that the correction due to this is below the \%-level~\cite{dynamical2016}.

We already corrected for the valence strange quark mistuning on the ensembles C1, C2, M1 and M2 by simulating directly at the physical strange quark mass.
On the other ensembles (C0, M0, F1) we need to correct the strange quark mass by -1.1(5)\%, -4.8(7)\% and -0.6(8)\%, respectively. 
We define dimensionless coefficients $\alpha_\mc{O}$ which we assume to be independent of the light quark mass, but take into account their lattice spacing and heavy quark mass dependence.
They are defined as the slope of an observable with the strange quark mass,
\eq{
  \mc{O}^\mathrm{phys}(a,am_h) = \mc{O}^\mathrm{uni}(a,am_h) \left(1 +\alpha_\mc{O}(a,am_h) \frac{m_s^\mathrm{phys}-m_s^\mathrm{sea}}{m_s^\mathrm{phys}} \right).
}
Having found the coefficients $\alpha_\mc{O}$ for the relevant observables $\mc{O}$ we can correct C0 and M0 directly.
For F1 we first interpolate to the corresponding values of the lattice spacing and the heavy quark mass, set by $m_{\eta_c}$.

\begin{figure}
  \begin{center}
    \includegraphics[width=0.49\textwidth]{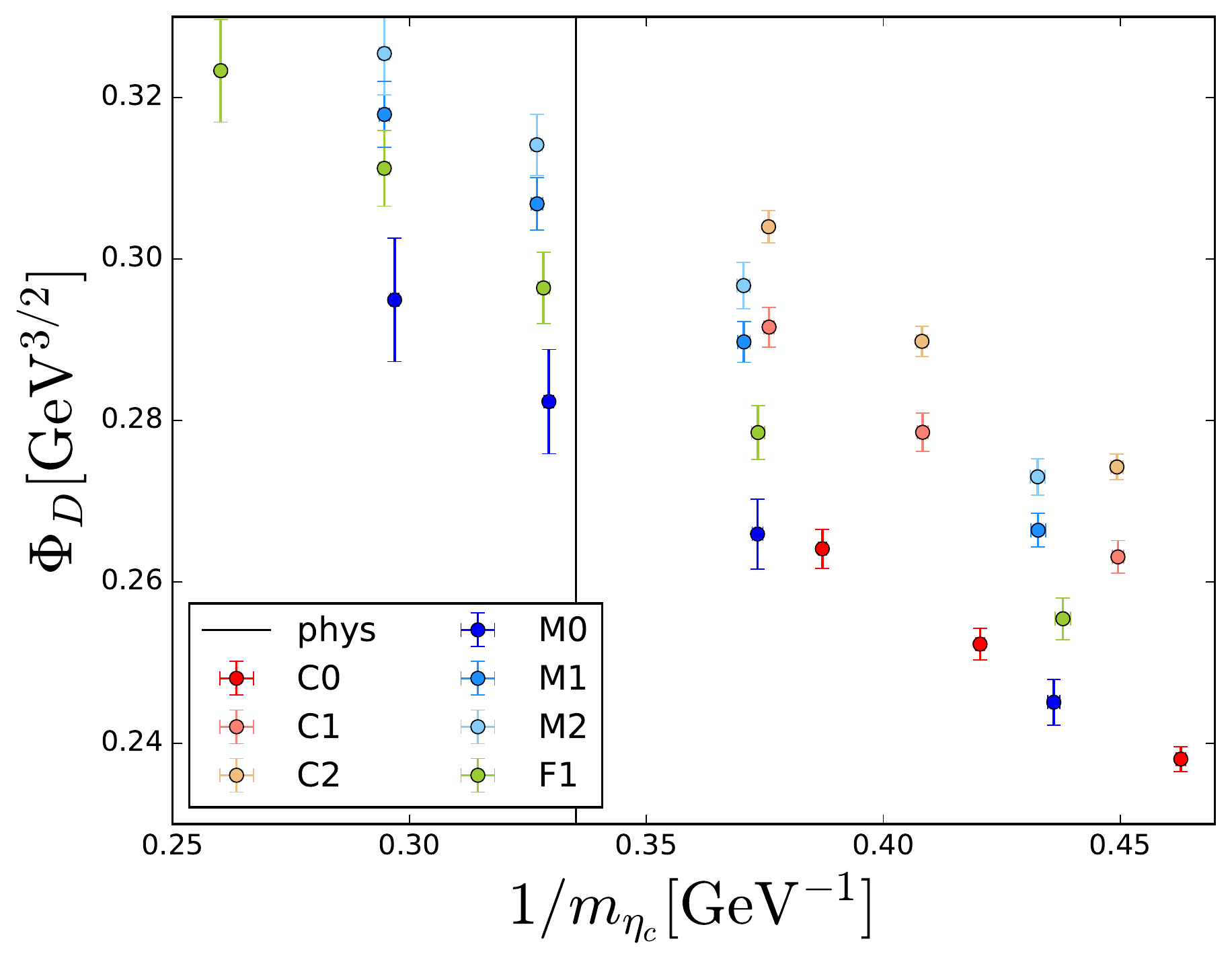}
    \includegraphics[width=0.49\textwidth]{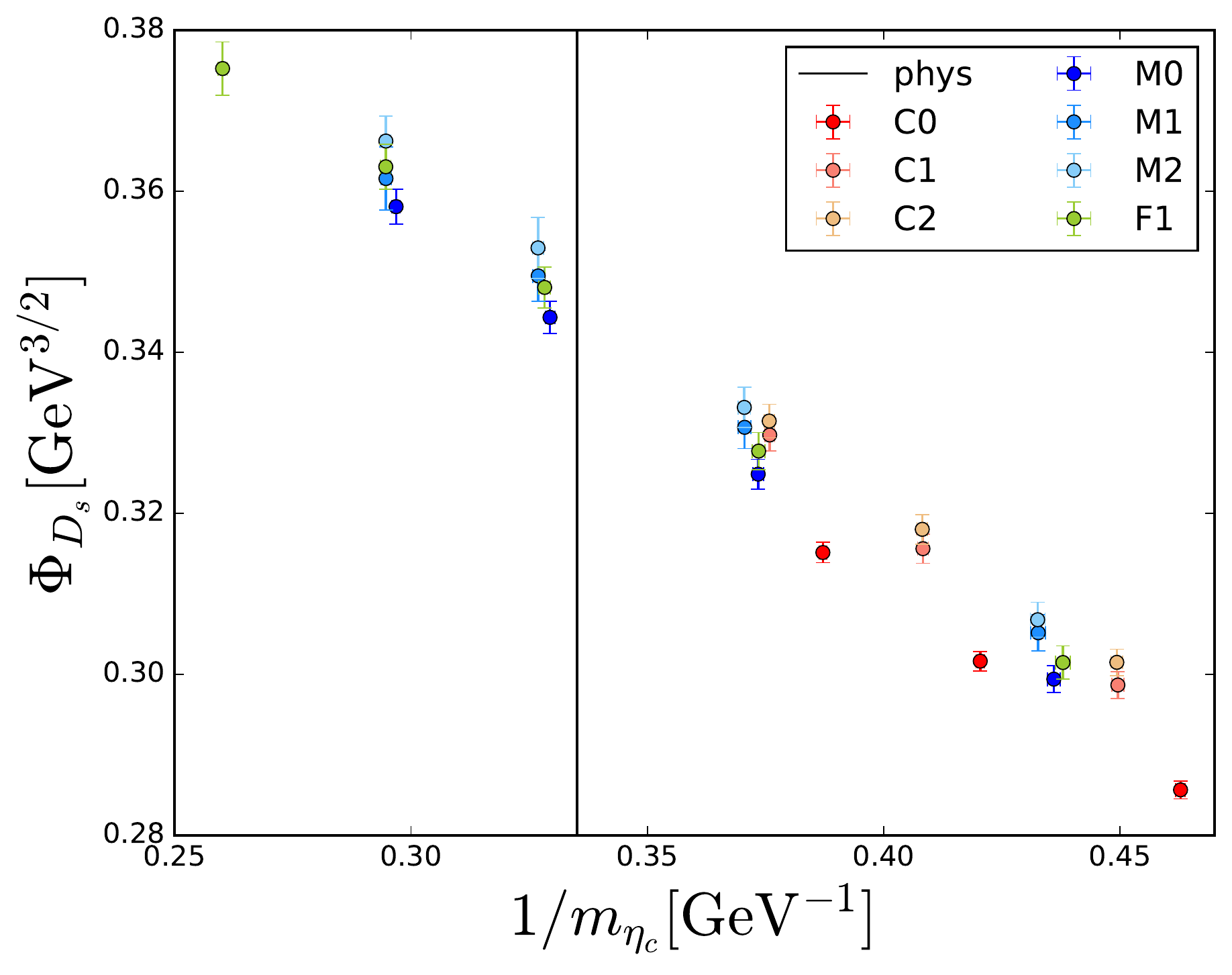}
  \end{center}
  \caption{$\Phi_D$ (left) and $\Phi_{D_s}$ (right) as a function of $1/m_{\eta_c}$. The red (blue, green) data points correspond to the coarse (medium, fine) ensembles. The vertical black line indicated the experimental value of $m_{\eta_c}$.}
  \label{fig:data}
\end{figure}

A first impression of the renormalised data at the physical strange quark mass is shown in Figure \ref{fig:data}. We notice the linear behaviour in the inverse heavy quark mass and the fact that the bound $am_h\lesssim 0.4$ prevents us from reaching the physical charm quark mass on the coarse ensembles.

\begin{figure}
  \begin{center}
    \includegraphics[width=0.495\textwidth]{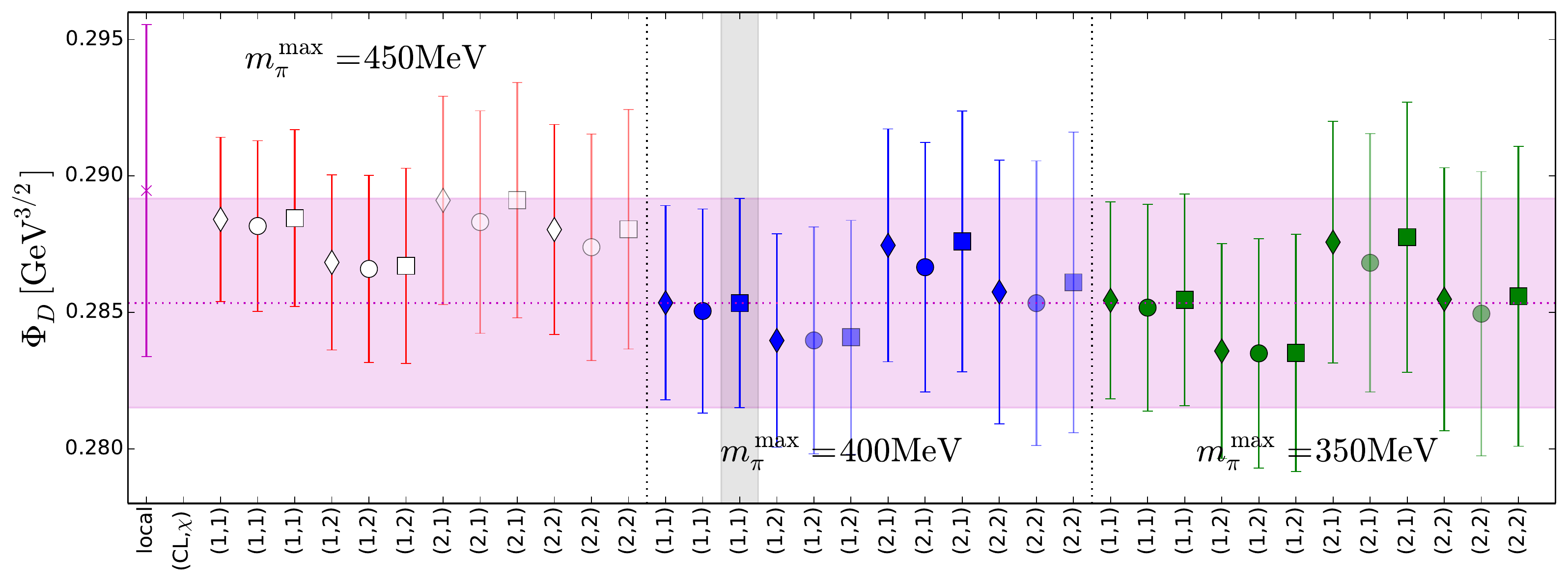}
    \includegraphics[width=0.495\textwidth]{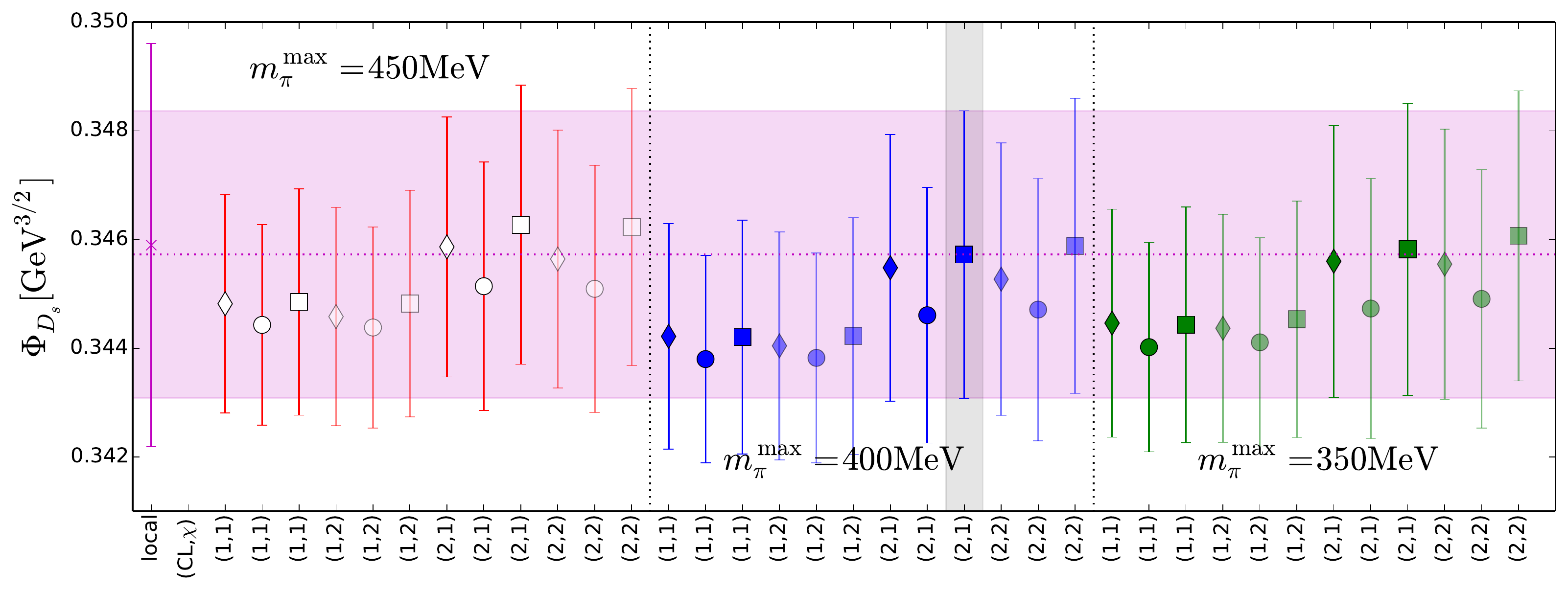}
  \end{center}
  \caption{Comparison of the different choices in the global fit. The grey and magenta bands highlight the preferred fit. The different symbols indicate different ways of fixing the heavy quark mass, i.e. $H=\,$ $D(\Diamond),\, D_s (\ocircle),$ and $\eta_c^\mathrm{connected}(\square)$. Fainter data points indicate that at least one of the heavy mass dependent coefficients is compatible with zero at the one sigma level. The label on the $x$-axis describes the number of coefficients for the continuum limit (CL) and pion mass limit ($\chi$), respectively. E.g. a fit labelled (2, 1) correspond to keeping two coefficients for the continuum limit extrapolation, but setting $C^1_\chi$ to zero. The magenta data point at the very left is the result of the analysis as obtained by the method presented in ref \cite{Boyle:2015kyy}.}
  \label{fig:PhiDsys}
\end{figure}

In the final step we extrapolate to $a=0$ and mesons corresponding to physical light and heavy quark masses. We use an expansion around the physical values given by
\eq{
  \mc{O}(a,m_\pi,m_H) = \mc{O}(0,m_\pi^\mathrm{phys},m_H^\mathrm{phys}) + \left[C_{CL}^0 + C_{CL}^1 \, \Delta m_H^{-1} \right] a^2 + \left[C_{\chi}^0 + C_{\chi}^1 \, \Delta m_H^{-1} \right] \Delta m_\pi^2 + C_h^0 \, \Delta m_H^{-1},
  \label{eq:globalchiCL}
}
where $\Delta m_H^{-1} = 1/m_H - 1/m_H^\mathrm{phys}$, $\Delta m_\pi^2=m_\pi^2 - {m_\pi^{2}}^{\mathrm{phys}}$ and $H = D, D_s$ or $\eta_c$.
We carried out this fit with different pion mass cuts ($m_\pi^\mathrm{max}=350, 400, 450\, \mathrm{MeV}$), different ways of setting the heavy quark mass ($H = D, D_s, \eta_c$) and the variations of whether or not mass dependent discretisation and physical pion mass extrapolations ($C^1_{CL}$ and $C^1_\chi$) are included.
Figure \ref{fig:PhiDsys} shows these variations for $\Phi_D$ (left) and $\Phi_{D_s}$ (right). 

The spread of the different results will be used to estimate the systematic error attached to the choice of fit ansatz. We postpone a detailed discussion of all systematic errors (which are at most as large as the statistical error), to a forthcoming publication~\cite{dynamical2016}. For now we simply state the statistical error for the preferred fit ansatz in both cases
\al{
   \Phi_D = 0.2853(38)_\mathrm{stat}\,\mathrm{GeV}^{3/2} \qquad &\Rightarrow \qquad f_{D} = 208.7(2.8)_\mathrm{stat}\,\mathrm{MeV}\\
   \Phi_{D_s} = 0.3457(26)_\mathrm{stat}\,\mathrm{GeV}^{3/2} \qquad &\Rightarrow \qquad f_{D_s} = 246.4(1.9)_\mathrm{stat}\,\mathrm{MeV}.\\
}

%%%%%%%%%%%%%%%%%%%%%%%%%%%%%%%%%%%%%%%%%5
\section{Outlook and Conclusions} \label{sec:conclusions}
\begin{figure}
  \begin{center}
    \includegraphics[width=0.49\textwidth]{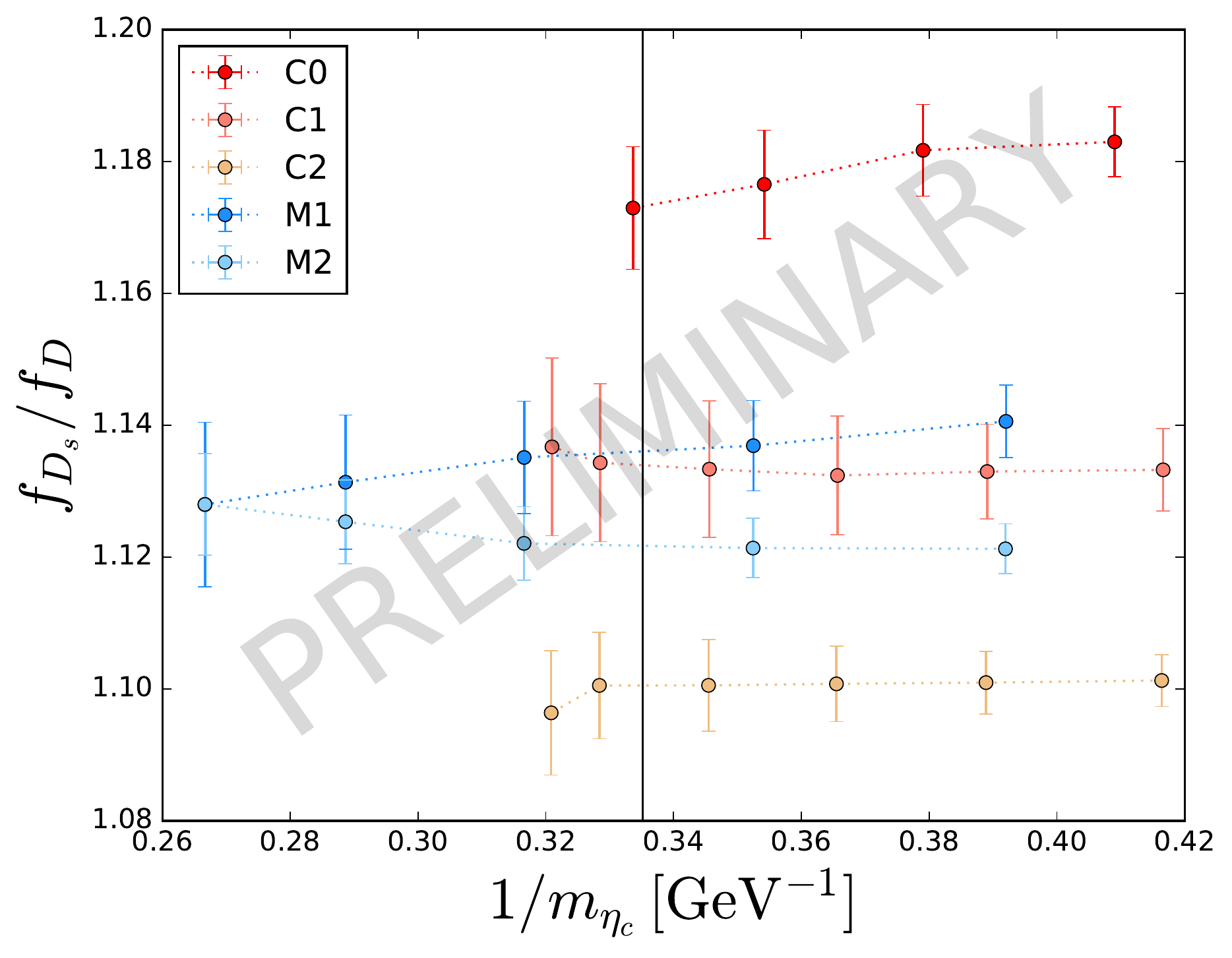}
  \end{center}
  \caption{First results for the ratio of decay constants $f_{D_s}/f_D$. Here the light and strange quarks use the unitary action, whilst the heavy quark is produced from a stout link smeared gauge field with a domain wall height of $M_5=1.0$. Similarly to the above, the solid black line corresponds to the physical value of $m_{\eta_c}$ and only serves to guid the eye.}
  \label{fig:smeared}
\end{figure}

%%%%%%%%%%%%%%%%%%%%%%%%%%%%%%%%%%%%%%%%%5
Following the strategy outlined at the end of ref \cite{Boyle:2015kyy} we are currently exploring how to extend the reach in the heavy quark mass by employing gauge link smearing. We find that with three hits of stout smearing with the standard parameter $\rho=0.1$~\cite{Morningstar:2003gk} and the choice $M_5=1.0$ we are able to extend the reach in the bare input quark mass up to $am_h \sim 0.69$. This is in agreement with findings of the JLQCD collaboration~\cite{Kaneko:2013jla,Hashimoto:2014gta}. With this we are able to simulate the physical charm quark mass even on the coarsest ensemble as can be seen from our first results presented in Figure \ref{fig:smeared}.

We presented our strategy to obtain decay constants $f_D$ and $f_{D_s}$ from $N_f=2+1$ flavour domain wall fermion ensembles with three lattice spacings in the range $a^{-1}=1.73-2.77\,\mathrm{GeV}$ and physical pion masses. We perform several different variations of the fit to obtain results at physical quark masses and in the continuum.

Furthermore we outline our current strategy to reach further in the heavy quark mass, allowing to simulate directly at the physical charm quark mass even on the coarse ensembles. First results look promising for the calculation of $D$ and $D_s$ phenomenology, such as semi-leptonic decays, and extrapolations of decay constants and other observables from the charm mass region to the bottom mass.

\clearpage
\acknowledgments{The research leading to these results has received funding from the European Research Council under the European Union's Seventh Framework Programme (FP7/2007-2013) / ERC Grant agreement 279757,  the Marie Sk{\l}odowska-Curie grant agreement No 659322, the SUPA student prize scheme, Edinburgh Global Research Scholarship and STFC, grants ST/M006530/1, ST/L000458/1, ST/K005790/1, and ST/K005804/1, ST/L000458/1, and the Royal Society, Wolfson Research Merit Award, grants WM140078 and WM160035 and the Alan Turing Institute. The authors gratefully acknowledge computing time granted through the STFC funded DiRAC facility (grants ST/K005790/1, ST/K005804/1, ST/K000411/1, ST/H008845/1). 
%%%%%%%%%%%%%%%%%%%%%%%%%%%%%%%%%%%%%%%%%5

  \bibliographystyle{JHEP}
  \bibliography{lattice2016.bib}

\end{document}